\documentclass[aps,prr,twocolumn,
superscriptaddress
]{revtex4-2}
\usepackage{amsfonts}
\usepackage{amsmath}
\usepackage{mathrsfs}
\usepackage{amssymb}
\usepackage{graphicx}
\usepackage{lipsum} 
\usepackage[T1]{fontenc}
\usepackage{epstopdf}
\usepackage{bm}
\usepackage{color}

\usepackage[colorlinks=true,linkcolor=blue,anchorcolor=blue,citecolor=blue,urlcolor=blue]{hyperref}

\newcommand{\cN}{{\cal N}}

\begin{document}

\title{Families of localized modes of Bose-Einstein condensates enabled by incommensurate optical lattice and photon-atom interactions}

\author{Pedro S. Gil}
\affiliation{
Departamento de F\'isica and Centro de F\'isica Te\'orica e Computacional, Faculdade de Ci\^encias, Universidade de Lisboa, Campo Grande, Edif\'icio C8, Lisboa 1749-016, Portugal }
\author{Vladimir V. Konotop}
\email{vvkonotop@ciencias.ulisboa.pt}
\affiliation{
	Departamento de F\'isica and Centro de F\'isica Te\'orica e Computacional, Faculdade de Ci\^encias, Universidade de Lisboa, Campo Grande, Edif\'icio C8, Lisboa 1749-016, Portugal 
}

\begin{abstract}
 
We consider a Bose–Einstein condensate (BEC) loaded into a one-dimensional optical cavity under the combined action of an external potential and atom-cavity coupling with mutually incommensurate periods. Such configuration enables the localization of matter waves even in the absence of two-body interactions. We study families of localized modes within the mean-field approximation for red and blue detunings from atomic and cavity resonances in relatively shallow quasi-periodic lattices, beyond the validity of the tight-binding approximation. The parameter regimes supporting localization of atomic wavepackets are identified. The system exhibits two types of bistability manifested as distinct photon numbers under otherwise identical conditions. One type arises from the coexistence of multiple families of localized modes, typical of conservative nonlinear systems, while the other stems from the multivalued dependence of the families on system parameters, characteristic of systems exhibiting hysteresis. BEC in a cavity may also display pseudo-degeneracy, understood as the existence of two distinct atomic density distributions corresponding to the same atomic and photon numbers (although different chemical potentials). The stability of the localized modes is analyzed. It is shown that, owing to the strong impact of long-range interactions on stability, a two-localized-mode configuration can operate as an XOR logic gate.
 
\end{abstract}

\maketitle

\section{Introduction}

Optical cavities containing cold atoms or Bose–Einstein condensates (BECs)~\cite{Haroche2006,Ritsch2013,Mivehvar2021} exhibit intrinsic nonlinearities and long-range interactions, leading to physical phenomena that have no counterparts in free space. Atoms in such systems experience two types of potentials with distinct physical origins: an external trapping potential, such as an optical lattice (OL) formed by interfering laser beams, and a cavity-induced potential arising from photon–atom interactions, which is also periodic but whose depth depends on the number of photons. When the total potential becomes incommensurate, i.e., when the ratio of periods of the OL and the cavity mode is irrational, localized one-dimensional (1D) atomic states can emerge~\cite{Zhou2011,Rojan2016,Major2018}. Incommensurability of backaction and a 2D external lattice enables existence of a Bose-glass phase~\cite{Habibian2013}.

Quasi-periodic 1D potentials for atoms can be created straightforwardly by using external  laser fields with incommensurate periods OLs~\cite{Roati2008,Deissler2010,Luschen2018,Wang2022}. In such potentials, which share several features with random ones, extended and localized states exist for different system parameters, as it is known due to celebrated  Aubry-Andr\'e (AA) model~\cite{Aubry1980}. This model provides a tight-binding approach to the bichromatic quasi-periodic lattices, when one constituent component is much deeper than the other. When the amplitudes of the two components are comparable, the conventional tight-binding approximation breaks down, and in the mean-field approximation the system must be described by the continuous Schr\"odinger equation with a quasi-periodic potential. In this regime, localized and extended modes may coexist at the same values of the system parameters but in different energy domains,  as proven mathematically in~\cite{Frohlich1990,Surace1990} and studied numerically in the physical literature over (see, e.g.~\cite{Diener2001,Modugnho2009,Biddle2010,Prates2022} and references therein).  Note, however, that in shallow quasi-periodic potentials bearing localized states it is still possible to use a basis of localized states reducing description of the system to nonlinearly coupled discrete lattices~\cite{Konotop2024}. 

The quantum potential possesses a distinctive property: since it originates from the interaction between atoms and photons, it depends both on the total number atoms and on their spatial distribution in the cavity, being inherently nonlocal~\cite{Ritsch2013,Mivehvar2021}. Consequently, when quasi-periodicity arises from the incommensurate wavelengths of the optical lattice (OL) and cavity mode, any eventual localization is inherently nonlinear. This nonlinearity stems from photon–atom interactions and can therefore manifest itself even in cavities containing only a few atoms. For example, one-dimensional localization of a single atom has been reported in~\cite{Rojan2016,Major2018}. We also note, that single particle localization in a quasi-periodic bichromatic lattice, without coupling to a cavity field and unaffected by two-body interactions, was observed experimentally in \cite{Roati2008,Luschen2018}.

The emergence of localized states of a BEC in a cavity with incommensurate 1D external and quantum potentials was analyzed in~\cite{Zhou2011} within the tight-binding approximation governed by the AA model. It was shown that the mismatch between the OL wavelength and the cavity mode induces a localization–delocalization transition. Moreover, the system exhibits bistability of the cavity field, involving localized and extended atomic states. More generally, bistability is an inherent property of atomic systems in cavities and does not necessarily rely on incommensurability, as demonstrated experimentally for a Fabry-Perot cavity with rubidium atoms~\cite{Gupta2007}, studied  theoretically in~\cite{Kazemi2016} within the two-mode approximation, and in~\cite{Diver2014} for a continuous model including inter-atomic interactions.    

Intrinsic nonlinearity of the photon-atom interactions can be interpreted as a form of many-body localization (see e.g.~\cite{Abanin2019} for a review). This aspect for incommensurate potentials in cavities was explored in studies of the effect of quasiperiodicity and nonlocality on ergodicity breaking~\cite{Kubala2021}. 
Other effects, like, superradiant instability of bosons in a cavity subject to quasi-periodic potentials, was also demonstrated recently~\cite{Yin2020}.
 
While localization induced by the combined action of an incommensurate OL and cavity backaction is well established, previous studies have mainly addressed particular stationary solutions, leaving many nonlinear phenomena unexplored. These effects, which constitute the main objective of the present paper, originate from the coupling between a dissipative cavity field and a condensate with a conserved number of atoms. First, we show that non-interacting Bose–Einstein condensates (BECs) in optical cavities support families of nonlinear modes characterized by a dependence of the atom number on the chemical potential and, consequently, on the cavity field intensity. This contrasts with conservative BECs loaded into OLs, where families of nonlinear modes (solitons) exist only in the presence of two-body interactions. Furthermore, the reported families may exhibit either one or two nonlinear mobility edges (MEs), separating localized and extended states within the same branch. In particular, depending on the atomic and cavity detunings, an increase in the number of atoms may lead either to localization or to delocalization of the atomic states. We also report the coexistence of multiple families and their multivalued character, resulting in two distinct types of bi- or multistability. Finally, the prominent role of long-range interactions allows the observation of diverse dynamical regimes, some of which are analogous to the operation of logic gates, specifically the XOR gate.

The organization of the paper is as follows. The model is formulated in Sec.~\ref{sec:model}. Families of localized modes are described in Sec.~\ref{sec:loclaization} where different types of the probe field detuning from atom and cavity resonances are discussed.  In Sec.~\ref{sec:multistability} we address the bistability of such systems. Excitation and dynamics of stable and unstable localized modes are described in Sec.~\ref{sec:dynamics}. The outcomes are summarized in the Conclusion.
 
\section{The model}
\label{sec:model}

The simplest dimensionless equations governing a noninteracting BEC coupled with cavity photons in the mean-field approximation read~\cite{Brennecke2008,Nagy2008,Ritsch2013,Mivehvar2021} 
\begin{align}
	\label{eq:Psi}
	i\frac{\partial \Psi}{\partial t}&=-\frac{1}{2}\frac{\partial^2\Psi}{\partial x^2}+[V(x)+|\alpha|^2U(x)]\Psi, 
	\\
	\label{eq:alpha}
	i\frac{d \alpha}{d t}&=- \Delta\alpha+i(\eta-\kappa\alpha)+\alpha\int   U(x)|\Psi(x,t)|^2dx .
\end{align}
Here $\Psi(x,t)$ is the macroscopic wavefunction of the BEC and $\alpha$ is the complex amplitude of the cavity field. The spatial coordinate is measured in the units of $\lambda_c/2\pi$ where $\lambda_c$ is the wavelength of the cavity mode creating $\pi$-periodic lattice $U(x)=U_0\cos^2x$, while time is measured in the units of $\omega_r^{-1}=\hbar/(2E_r)$ where $E_r=h^2/(2m\lambda_c^2)$ is the recoil energy associated with the cavity mode. The single-photon lattice depth normalized to the recoil frequency is given by $U_0=g^2/(\Delta_a\omega_r)$ where $g$ is the strength of the atom-cavity coupling and $\Delta_a=\omega-\omega_a$ is the detuning between the pump-laser and the atomic transition frequencies. In the chosen units $\Delta=\Delta_c/\omega_r$, with $\Delta_c=\omega-\omega_c$, is dimensionless cavity detuning. The rate $\eta$ of the laser coherently driving the cavity and the cavity losses $\kappa$ are measured in the $\omega_r$ units.

We assume that the external potential created by laser beams has a spatial period $\lambda_c/2\beta$ where $\beta$ is an irrational number. Respectively, in the dimensionless units (energy is measured in the units $2E_r$) we set $V(x)=V_0\cos^2(\beta x +\vartheta)$, where an arbitrary phase shift $\vartheta$ is introduced to break the parity symmetry (physical effects with BEC in even quasi-periodic potentials were discussed in~\cite{Prates2022}) and  $V_0$ is the lattice depth measured in the units of the recoil energy $E_r$.

Equation~(\ref{eq:Psi}) will be considered subject to the zero boundary conditions, $\Psi(\pm L/2,t)=0$ where $L$ is the dimensionless condensate length, that also defines the integration domain in (\ref{eq:alpha}). Such setting models a BEC loaded in a sufficiently deep~\cite{Jaouadi2010,Gaunt2013,Navon2021} and long 1D potential box. Most importantly, in this work we focus on localized states that are centered sufficiently far from the boundaries and are therefore largely unaffected by the specific boundary conditions, unlike extended modes. Consequently, the formal limit $L\to\infty$ can also be considered.
 
The macroscopic wavefunction $\Psi(x,t)$ is normalized to the total number of atoms $N$: $\int|\Psi|^2dx=N$. While we  neglected two-body interactions in the GPE (\ref{eq:Psi}) 
the number of atoms still determines the shift of the cavity resonance, given by the integral $\int U(x)|\Psi|^2dx$ and consequently the amplitude of the back action $U(x)$. In other words, it remains the main parameter characterizing the families of the nonlinear modes which are discussed in detail in the next Sec.~\ref{sec:loclaization}.

The dimensionless system (\ref{eq:Psi}), (\ref{eq:alpha}) involves several control parameters. To reduce their number while analyzing the solutions, we fix the absolute values of the cavity and atomic detunings, $|\Delta_c|$ and $|\Delta_a|$, as well as cavity losses $\kappa$, while allowing the driving amplitude $\eta$ to vary.   This allows one to scale out $|U_0|$ by the anstaz $\tilde{\alpha}=|U_0|^{1/2}\alpha$ and  redefine the driving parameter $\tilde \eta=|U_0|^{1/2}\eta$. Introducing also
\begin{align}
	\label{Theta}
	 \Theta(t)=\mathcal{N}\int_0^td\tau \int dx \cos^2(x)|\Phi(x,\tau)|^2,
\end{align}
where $\mathcal N=|U_0|N$ is the re-scaled number of atoms, as well as $\Phi=\Psi/\sqrt{N}$, ensuring the normalization $\int|\Phi|^2dx=1$,
Eq.~(\ref{eq:alpha}) is readily solved: 
\begin{align}
	\label{t_alpha}
	 \tilde{\alpha}=\left[\tilde\eta \int_{0}^{t}dt'
	 e^{-(i\Delta-\kappa)t'+i\sigma\Theta(t')}
	 +\tilde{\alpha}_0\right]e^{(i\Delta-\kappa)t-i\sigma\Theta(t)}
\end{align}
where $\sigma=$sign$\Delta_a$=sign$U_0$, and $\tilde\alpha_0$ is the initial amplitude of the photon field (in the chosen units). Equation~(\ref{eq:Psi}) now acquires the form
\begin{align}
	\label{eq:Psi_1}
	i\frac{\partial \Phi}{\partial t}=-\frac{1}{2}\frac{\partial^2\Phi}{\partial x^2}+[V_0\cos^2(\beta x +\vartheta)+\sigma|\tilde \alpha|^2\cos^2x]\Phi 
\end{align}

Stationary solutions of (\ref{eq:Psi}), (\ref{eq:alpha}) are obtained by the anstaz $\Phi(x,t)=e^{-i\mu t}\phi(x)$ and $\tilde\alpha(t)=\tilde\alpha_{st}$. Now Eq.~(\ref{eq:Psi}) is reduced to the eigenvalue problem
\begin{align}
	\label{eigen}
	\mu{\phi}= -\frac{1}{2}\frac{d^2\phi}{dx^2}+\left[V_0\cos^2(\beta x +\vartheta)+\sigma A\cos^2x\right]\phi
\end{align}
where $A=A[\phi]=|\tilde\alpha_{st}|^2$ is the number of cavity photons (intensity of the cavity mode). It is a functional of the atomic wavefunction: from (\ref{eq:alpha}), it is defined by the well-known formula
\begin{align}
	\label{A}
	A=\frac{\tilde\eta^2}{\left(\sigma\Delta-\theta\right)^2+\kappa^2} 
\end{align}
where  	
	$\theta = \mathcal N 	\langle \cos^2x\rangle$ and the angular brackets are used for the spatial averaging averaging: $\langle f\rangle\equiv \int f(x) |\phi|^2dx$.

All possible atomic distributions are solutions of the linear eigenvalue problem (\ref{eigen}), i.e., they have the forms of the linear modes of the Schr\"odinger equation (\ref{eigen}) with the incommensurate bichromatic potential. It is known that they can be either localized or extended~\cite{Frohlich1990,Surace1990,Diener2001,Modugnho2009,Biddle2010,Prates2022,Konotop2024}, which depend on the amplitude, $V_0$, of the external potential, and on the amplitudes $\sigma A$ of the cavity field.
However, unlike in the mentioned previous studies, the peculiarity of the problem at hand is that the strength of the back-action depends on the atomic distribution, even for a fixed number of atoms.


Several general conclusions follow directly from (\ref{eigen}) and (\ref{A}). The presence of both periodic lattices of finite amplitudes is necessary for existence of localized solutions of (\ref{eigen}). Meantime the number of photons cannot exceed  $A_{\rm max}$ where
\begin{align}
	\label{Amax}
	A_{\rm max}=\begin{cases}
		 \displaystyle{\frac{\tilde\eta^2}{\kappa^2}}, &\sigma\Delta>0
		 \\
		 \displaystyle{\frac{\tilde\eta^2}{\Delta^2+\kappa^2}}, &\sigma\Delta<0
	\end{cases} 
\end{align}
Let us consider the formal limit $\mathcal{N}\to0$. From a physical point of view, this limit cannot be directly applied to single-atom localization, as such a case requires a fully quantum treatment, whereas Eqs.~(\ref{eq:Psi}) and (\ref{eq:alpha}) are derived within the mean-field approximation.
However, from the viewpoint of dynamical systems, this analysis allows one to establish critical system parameters and number of atoms when the localization may occur.

At $\sigma\Delta<0$ and $\cN=0$, localized modes can exist only if the driving amplitude exceeds a critical value $\tilde\eta_{\rm LDT}$ at the localization-delocalization transition occurs meaning that for weaker driving $\tilde \eta< \tilde\eta_{\rm LDT}$ all atomic estates are extended. In this case, an increase in the atomic density, leading to  increase in $\theta$, promotes delocalization, as it reduces the number of photons. Conversely, when the driving amplitude is sufficiently large, $\tilde{\eta} > \tilde{\eta}_{\rm LDT}$, localized modes exist already in the linear limit but become delocalized as $\mathcal{N}$ increases. One can thus identify a {\em nonlinear ME}, $\mathcal{N}_{\rm ME}$, above which no localized states are found.  

The existence of the nonlinear ME (regardless of the sign of $\sigma\Delta$) stems form the following more rigorous arguments. There exits a coordinate $x_*\in[0,\pi/2]$, such that $ \theta= \mathcal{N} \cos^2x_* $. For smooth (differentiable) distribution $|\phi|^2$ we have that $x_*\neq\pi/2$, because otherwise $\theta=0$, which is impossible for an integral of the product of two non-negative differentiable functions.
Thus, $\theta\to \infty$ when $\mathcal{N}\to\infty$ and consequently $A\to 0$.
This means that for  $\cN>\mathcal{N}_{\rm ME}$ (other parameters being fixed) all atomic states are extended.

While the described scenario can be realized also for $\sigma\Delta>0$, with smaller $\tilde{\eta}_{\rm LDT}$, now  even if all states are delocalized at $\cN=0$, 
when $A=A_0=\eta^2/(\Delta^2+\kappa^2)$ localization may be induced by increase of the atomic density, if $A=A_{\rm max}$ enables localization. In this case one can define the lower,  $\mathcal{N}_{\rm ME}^l$, and upper, $\mathcal{N}_{\rm ME}^u$ nonlinear mobility edges, such that the localized states exist for 
$\mathcal{N}_{\rm ME}^l <\mathcal N<\mathcal{N}_{\rm ME}^u$ and do not exist outside this interval.

As a concluding remark, the localization-delocalization transition $\tilde{\eta}_{\rm LDT}$ is a property of the entire system, whereas the nonlinear mobility edge $\mathcal{N}_{\rm ME}$ (or edges $\mathcal{N}_{\rm ME}^{l,u}$) is a property of individual families of solutions, in the sense that its value may differ from one family to another. This is illustrated below using numerical examples.
	
\section{Localized modes and their families}
\label{sec:loclaization}

Families of solutions parameterized by a control parameter provide an overview of the system as a whole, whereas specific examples of localized modes depend on the chosen number of atoms, density distributions, number of coupled cavity photons, etc. Here we address both aspects. Since the system under consideration consists of a conservative subsystem (the BEC) coupled to a dissipative one (the photons), the families of solutions can be represented either by curves $\cN(\mu)$, as is customary for conservative systems, or by curves $A(\cN)$, useful for identifying bistability domains [where, for a given number of atoms  $\cN$ two distinct values $A_1(\cN)\neq A_2(\cN)$ coexist]. The latter parametrization also allows one to determine the number of atoms that ensures the maximum number of cavity photons that can couple to localized modes. Either parametrization makes it possible to establish the maximum number of atoms supported by the cavity modes and to delineate the domains of dynamical stability of the solutions.  

The families shown in all figures below are numbered according to the increasing order of the chemical potentials computed at $\cN=0$ (regardless whether modes are extended in this limit, like in Figs.~\ref{fig1} and ~\ref{fig3},  or localized like in Fig.~\ref{fig4}). In all figures we show the lowest 88 families which possess intervals of localization, i.e. featuring nonlinear MEs; all higher families (checked numerically) correspond to extended modes only (not shown), in accordance with the discussion in previous section.   Thus, $\mu_1(\cN=0)\leq\mu_2(\cN=0)\leq\dots\leq\mu_{88}(\cN=0)$; note that the ordering of $\mu_j(\cN)$ with respect to their magnitudes at $\cN>0$ may change due to crossings between families. However, this does not affect the numbering of the families. To characterize localization of the modes we use the inverse participation ratio (IPR): $\chi= \int|\phi(x)|^4 dx$ (recall that the modes $\phi$ are normalized to one). 

Regarding the physical parameter regime for which the present analysis is applicable, we consider $^{87}$Rb atoms interacting with a cavity mode of wavelength $\lambda_c \approx 780\,\text{nm}$. The corresponding recoil frequency is estimated as $\omega_r \approx 47\,\text{kHz}$. Taking into account that the coupling constant $g$ can vary in the range of $1$–$100\,\text{MHz}$ and that the atomic detuning $\Delta_a \sim 217$–$619\,\text{GHz}$ for a pump wavelength $\lambda_p \sim 780$–$780.13\,\text{nm}$ (see, e.g.,~\cite{Ottl2006,Baumann2010,Steck2025}), we estimate that $|U_0| \lesssim 1$. The cavity detuning $\Delta_c$ is highly tunable (see, e.g.,~\cite{Kollar2015}), and values up to $|\Delta| \lesssim 500$ can be considered, while the cavity losses can be estimated as $\kappa \lesssim 100$ (e.g., in \cite{Kollar2015}  $\kappa\sim18$, or $\sim0.8\,\text{MHz}$ while in \cite{Kessler2014} $\kappa\sim0.6$, or $\sim28\,\text{kHz}$).

Accordingly, in the simulations reported below, we use $|U_0| = 0.5$, $V_0 = 0.5$ (both are $\sim23.5\,\text{kHz}$, in physical units), $|\Delta| = 150$ ($\sim7\,\text{MHz}$, in physical units), and $\kappa = 63$ ($\sim3\,\text{MHz}$, in physical units). The incommensurability of the periods is chosen in a form of the golden ratio, $\beta=(1+\sqrt{5})/2$. This choice, however, is not essential for existence of localized modes: any other irrational number can be used as well. Moreover, it can be approximated by one of its best rational approximations, without any significant impact on the results (see e.g., discussion in~\cite{Konotop2024}). One of these approximations is given by $\beta\approx 144/89$. In that case, the system can be (formally) periodically extended over the whole real axis $x$ with the period $89\pi$. This consideration determines our choice of the
dimensionless condensate length as fixed at $L = 89\pi$, which corresponds to $34.71\,\mu\text{m}$ for the cavity mode chosen for simulations.  The constant $\vartheta$ does not qualitatively affect the results, provided it breaks inversion symmetry; in the simulations shown below it is arbitrarily chosen as $\vartheta=1.23$.

While the estimates presented in the preceding section are of general character, features of the localization depend qualitatively on the signs of the detuning from the atomic and cavity resonances. In particular, the sign of $\Delta_a$ (i.e.,  $\sigma$) determines the sign of the quantum potential in (\ref{eigen}). Therefore, below we consider four different cases, blue-blue ($\Delta_a,\Delta_c>0$), red-red ($\Delta_a, \Delta_c<0$), blue-red ($\Delta_a>0$, $\Delta_c<0$), and red-blue ($\Delta_a<0$, $\Delta_c>0$) detunings, separately.

\subsection{Blue-Blue detunings}

\begin{figure}
	\includegraphics[width=\columnwidth]{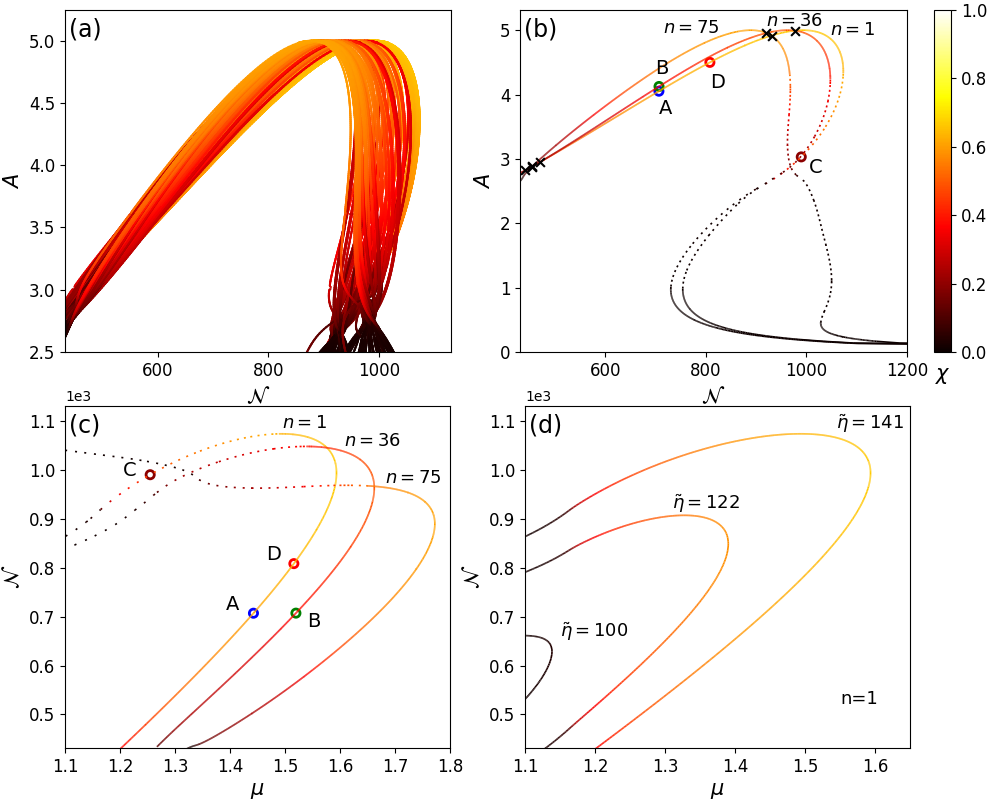}
	\caption{(a) Lowest 88 families of stationary modes for $\sigma=+1$ and $\Delta=150$ on the diagram $(A,\mathcal{N})$.  (b) Examples of three families from the bundle shown in (a) (note that the range of $A$ is changed). The crosses indicate crossing of the families. The colored circles, labeled "A" through "D", indicate specific modes which are used as examples below. (c) The families of the modes shown in (b) but now using the alternative parametrization  $\mathcal{N}$  {\it vs} $\mu$. In  (a), (b) and (c) $\tilde\eta=141$. In (b) and (c) solid and dashed lines correspond to intervals of dynamical stability and instability. (d) $\mathcal{N}(\mu)$ of the family $n=1$, for three different values of the driving amplitude  $\tilde\eta=100,122,141$. Here, and in all figures below, the degree of localization is encoded by the color scale indicated in the IPR color-bar.}
	\label{fig1}
\end{figure}

Starting with the case $\Delta_a,\Delta_c>0$ (hence $\sigma\Delta>0$), in Fig.~\ref{fig1}~(a) we present families of the stationary localized modes on the diagram $(A,\mathcal N)$ obtained for the fixed detunings, the cavity parameters specified above, and for the driving amplitude $\tilde\eta=141$, which is below the $\tilde{\eta}_{\rm LDT}\approx 270$ (while this value depends on the OL amplitude, it is independent of the detuning colors), i.e., all linear modes are extended. Only the intervals corresponding to localized modes ($\chi\gtrsim 0.1$), i.e., $\cN_{\rm ME}^l<\cN<\cN_{\rm ME}^u$ are shown entirely: outside this interval all modes are extended; see also Fig.~\ref{fig1}~(b). The positions of the nonlinear MEs are identified for each family by a relatively sharp change of the color corresponding to IPR values: the darker regions indicating larger spatial extension of states. We observe that regardless of mode localization, the number of photons is bounded from above  $A\leq A_{\rm max}$ (for $\tilde\eta=141$ we have $A_{\rm max} \approx 5$).  At the same time, for each family one can identify mobility edges $A_{\rm ME}^{l,u}$ such that at $A_{\rm max}\geq A>A_{\rm ME}^u$ all modes are localized while at $A<A_{\rm ME}^l$ all modes are extended (numerically $A_{\rm ME}^l$ e $A_{\rm ME}^u$ are very close to each other).   

The families shown in (a) are barely distinguishable on the scale of the figure. In Fig.~\ref{fig1} (b) we illustrate three well distinguishable ones. One observes that the families can cross with each other. At such points, marked by crosses, a pseudo-degeneracy occurs: the system with the same number of atoms and photons supports two different configurations (the respective localized modes of the condensate have different profiles and different chemical potentials). For all families the localization occurs around the highest intensities of the respective cavity mode.  In panel (b) we also show the stability of the modes which was determined by the direct propagation (stable and unstable modes are shown by solid and dashed lines respectively). To this end we solved numerically 
Eqs.~(\ref{eq:Psi}) and (\ref{eq:alpha}) with initially perturbed cavity intensity $\alpha(t=0)$ for times $ \gtrsim 19$. A mode was considered stable if $\alpha(t)$ converged to its stationary value, according to $ ( |\alpha(t)|^2-|\alpha_{\rm st}|^2 ) / |\alpha_{\rm st}|^2<0.005$ on the time-scale $1/\kappa\sim 0.01$, while oscillations of the position of the c.m. $X_j=\int x|\phi_j|^2dx$ were below $10^{-6}$ (note that change of the c.m. implies deviation of $|\alpha(t)|$ from its stationary value). 
 
Equation (\ref{eigen}) with $A$ replaced by its stationary value from (\ref{A}), represents stationary Schr\"odinger equations for atoms only, with a nonlocal nolinearity. For such conservative system a conventional parametrization of the families is given by the dependence $\mathcal{N}(\mu)$, which for the families from Fig.~\ref{fig1} (b), is illustrated in Fig.~\ref{fig1} (c). For a finite size cavity such families have the linear limit (not shown in the panel) where they are delocalized, i.e., they bifurcate from the linear spectrum.  

As it was predicted above, the localization is enabled by the long-range interactions and exhibits a lower nonlinear ME $\mathcal{N}_{\rm ME}^l$, i.e., localized modes of shown families exist only at $\mathcal{N}>\mathcal{N}_{\rm ME}^l$ (for the chosen parameters individual modes have $\mathcal{N}_{\rm ME}^l\approx 400$). Conversely, there is an upper nonlinear ME such that all states delocalize at $\mathcal N>\mathcal{N}_{\rm ME}^u$ (in the shown modes $\mathcal{N}_{\rm ME}^u\approx 1070$).  

The above results were obtained for a fixed driving strength. Localization or delocalization of atomic states can be efficiently controlled by varying the number of cavity photons. Figure~\ref{fig1}(d) shows the dependence of the lowest family ($n=1$) for three different values of $\tilde\eta$. Increasing the driving amplitude enhances localization and increases the interval  $\cN\in (\mathcal{N}_{\rm ME}^l,\mathcal{N}_{\rm ME}^u)$ where modes are localized.

\begin{figure}
	\includegraphics[width=\columnwidth]{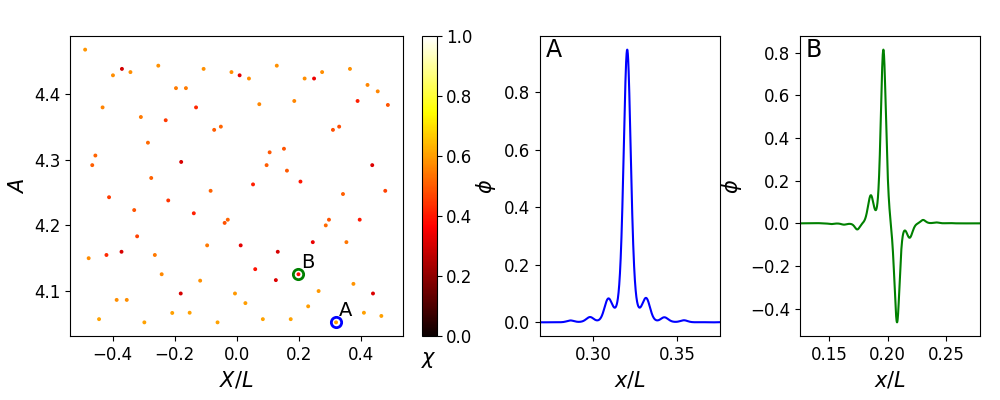}
	\caption{Distributions of the localized modes for $\sigma=+1$, $\Delta=150$ and $\tilde\eta=141$, with $\mathcal{N}\approx7\times10^2$ on the diagram $(X/L,A)$ (the leftmost panel). The panels on the right show the wavefunctions of the modes, labeled A and B in the left panel and in Figs. \ref{fig1} (b) and (c).}
	\label{fig2}
\end{figure}

Turning now to the spatial distribution of the localized modes in the cavity and the corresponding photon numbers, we present in Fig.~\ref{fig2} the diagram  $(X/L,A)$. It shows that the distribution of modes along the cavity length is nearly homogeneous, resembling the situation of BECs loaded into quasiperiodic potentials in the absence of photons~\cite{Prates2022,Konotop2024}. The distribution of localized modes (with their centers of mass indicated by dots) over the interval of cavity field intensities is nearly homogeneous: although two modes may occasionally be found at very close spatial positions, when considering intervals extending over several periods of either sublattice, the number of localized modes within such intervals is approximately the same, independent of the position of the chosen interval. Below, in Sec.~\ref{sec:dynamics} we provide an example of a possible applications of the mentioned distribution of the localized modes.

Modes that are spatially close may correspond to very different photon numbers. Conversely, modes associated with nearly the same photon number can be localized at widely separated positions along the cavity. This relation between the spatial localization of the modes and the cavity-field intensity implies that information about BEC localization in the cavity can be inferred from measurements of the cavity field.

\subsection{Red-Red detunings}

The families of nonlinear modes for other detuning configurations are simpler than those considered above for the blue–blue detuning. We now briefly discuss these other cases.

\begin{figure}
	\includegraphics[width=\columnwidth]{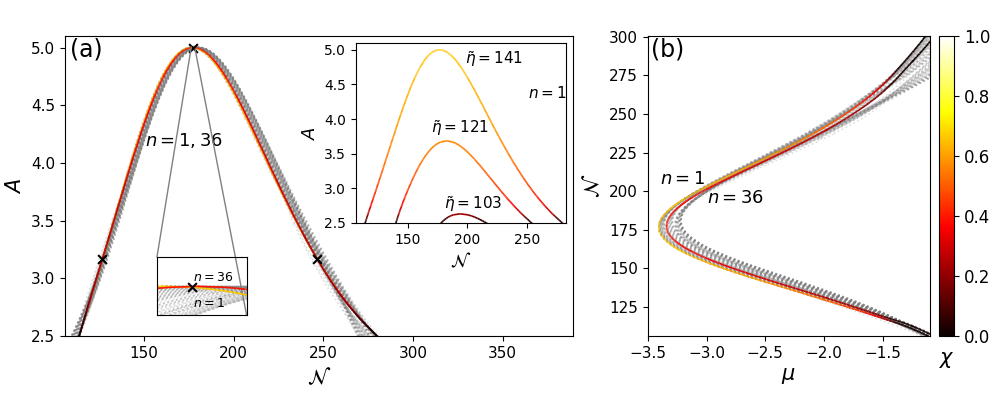}
	\caption{(a) Families of the cavity modes $A(\mathcal N)$ with intervals of localized solutions
		  for $\sigma=-1$, $\Delta=-150$, and $\tilde\eta=141$. Two families (discussed in the text) are highlighted by color specifying IPR, while the rest of the families are shown by light gray dashed lines. The right inset shows $A(\mathcal{N})$ for $n=1$ family for  $\tilde\eta=103,121,141$. The lower inset magnifies the intersection between families $n=1$ and $n=36$. (b) Number of atoms $\mathcal{N}$ {\it vs} chemical potential $\mu$ for the chosen families (in color, the rest of the families are shown by light-gray dashed lines). Solid (dashed) lines indicate stable (unstable) intervals.}
	\label{fig3}
\end{figure}

In Fig.~\ref{fig3} we show the families for the case of red-red detunings, $\Delta_a,\Delta_c<0$ (corresponding to $\sigma\Delta>0$). Unlike in the previous case, now the external lattice and  back-action potential are out-of-phase. We observe that $A(\mathcal{N})$ is a single-valued nonmonotonic function. Choosing  parameters with $\tilde\eta<\tilde\eta_{\rm LDT}$ but enabling localization at $A_{\rm max}$ we again obtain the MEs separating localized and delocalized modes. In the right inset in Fig.~\ref{fig3} (a) the family $n=1$ is shown for three different values of the driving amplitude $\tilde\eta$, illustrating the general behavior: increase of the external driving enhances the localization. Another generic feature is the crossing of the families $A(\mathcal{N})$, indicating the pseudo-degeneracy of the states. Recall that, unlike conventional bistability, where different modes with the same number of atoms are supported by different numbers of photons, at a pseudo-degeneracy point both the number of atoms and the number of photons are identical, and the differences between the degenerate solutions lies in their spatial atomic distributions and chemical potentials.

In Fig.~\ref{fig3} (b) we show the families on the diagram ($\mu,\cN$), where lower and upper nonlinear MEs are clearly seen. Unlike in the case of blue–blue detuning, the localized modes now appear at lower chemical potentials, while they become delocalized for $\mu>\mu_{\rm ME}$ [cf. Fig.~\ref{fig1} (b) and Fig.~\ref{fig1} (c)]. 

Another notable difference in the present case is that the number of atoms required for localization for red–red detuning is much smaller than that for blue–blue detuning [cf. Fig.~\ref{fig1}(a) and Fig.~\ref{fig3}(a)]. This effect is determined by the backaction. To understand it, let us consider the number of atoms in the modes corresponding to $A_{\rm max}$ defined in Eq.~(\ref{Amax}). In both cases this maximum is achieved at $ \cN\langle\cos^2(x)\rangle=\sigma\Delta$. In the present parameter regime, $A_{\rm max}$ is much larger than $U_0$; hence the spatial c.m. of the localized modes in both cases is governed by the backaction. Thus, it follows from Eq.~(\ref{eigen}), that atoms are concentrated in the vicinity of a minimum and a maximum of $\cos^2(x)$ for blue ($\sigma=1$) and red ($\sigma=-1$) atomic detunings, respectively. Consequently $\langle\cos^2(x)\rangle$ is much smaller in the former case, which requires a much larger $\cN$ to keep the product $\cN\langle\cos^2(x)\rangle$, and thus, $A_{\rm max}$, unchanged. 

\subsection{Detunings of different colors}
\label{sec:two_color}

\begin{figure}
	\includegraphics[width=\columnwidth]{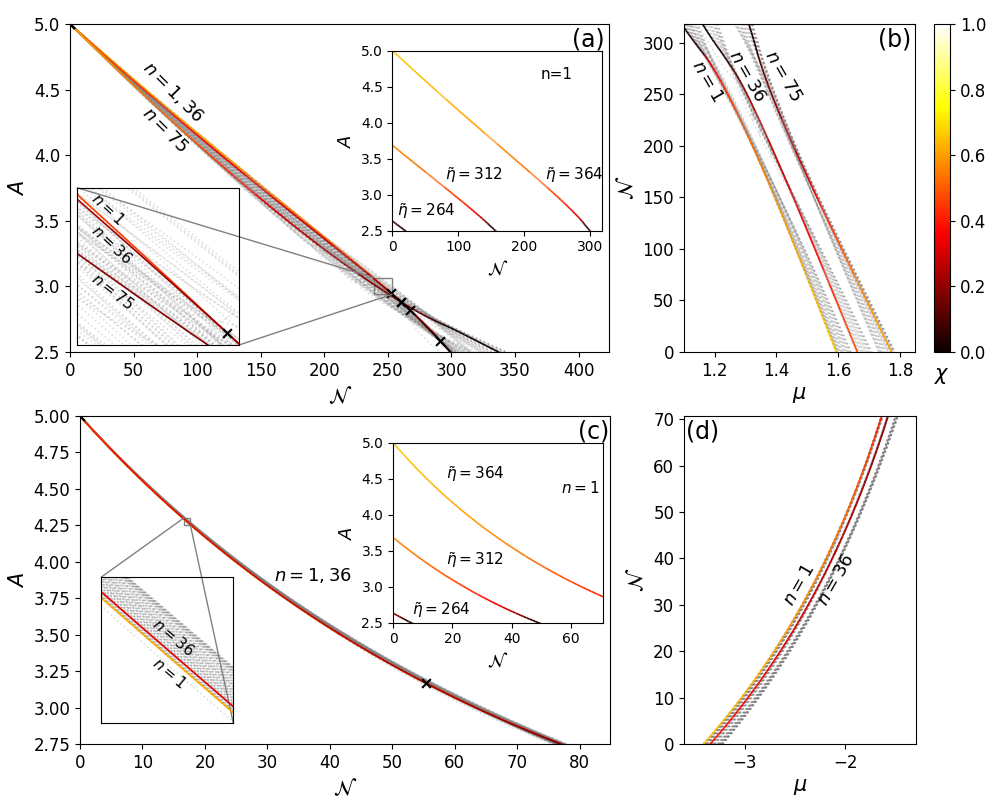}
	\caption{  Cavity field intensity $A$ {\it vs} the number of atoms $\mathcal{N}$
		for $\sigma=+1$ and $\Delta=-150$ in (a) and for  $\sigma=-1$ and $\Delta=150$ in (c). For clarity three families are highlighted by color indication of the IPR, while the remaining 85 families in each bundle are shown by gray color. In (a) and (c), $\tilde\eta=364$, and the bottom left insets show zoomed-in regions of their respective panel. In the insets in top right corner of (a) and (c), the respective family bifurcating from the lowest linear states, $n=1$, is shown for three different driving amplitudes, $\tilde\eta=264,312,364$.
        Panels (b) and (d) show the number of atoms $\cN$ {\it vs} the chemical potential $\mu$ for the same parameters and families shown in panels (a) and (c), respectively. Note that in (c) all families are very close to each other (two of them are zoomed in the left inset).
        In all panels $\kappa=63$.
		}
	\label{fig4}
\end{figure}

In Fig.~\ref{fig4}, we show families for the cases where cavity and atomic detunings have different colors: $\Delta_a>0$, $\Delta_c<0$  in panels (a), (b) and  $\Delta_a<0$, $\Delta_c>0$  in panels (c), (d). It follows from Eq.~(\ref{eq:alpha}), and is evident in Figs.~\ref{fig4}(a) and \ref{fig4}(c), that increasing the number of atoms is accompanied by a decrease in the number of photons coupled to the stationary modes. Consequently, if the pumping strength is below the LDT threshold $\tilde\eta_{\rm LDT}$, no localized atomic distributions can exist.  That is why, now we have chosen the driving amplitude above the localizing-delocalizing transition: $\tilde\eta=364>\tilde\eta_{\rm LDT}$. The localized atomic modes exist in the (formal) linear limit $\mathcal{N}=0$. Note that on the diagram ($A,\mathcal{N}$) all families start in the same point $(\cN=0, A_{\rm max})$, where $A_{\rm max}$ is defined in (\ref{Amax}) [obviously, this is also true for the families shown in Fig.~\ref{fig1}(a) and Fig.~\ref{fig3}(a)]

The  dependence $A(\mathcal{N})$ is now monotonic ($A\leq5$): increase of the number of atoms results in the decrease of the intensity of the cavity field. By comparing upper and lower panels one observes that the decay of $A$ with $\mathcal{N}$ is faster in the case of red detuning from the atomic resonance, which is explained by different signs of the external and quantum potentials in (\ref{eigen}): in this last case delocalization occurs at smaller number of atoms due to decrease of the depth of the combined bi-chromatic lattice. The difference between these two cases is more pronounced on the diagrams $(\mu,\cN)$ [cf. Fig.~\ref{fig4} (b) and Fig.~\ref{fig4} (d)] where the derivatives $d\cN/d\mu$ have opposite signs.

\section{Bistability}
\label{sec:multistability}

The modes discussed in the preceding section bifurcate from the linear ones, which can be either extended at $\tilde{\eta}<\tilde{\eta}_{\rm LDT}$ (Figs.~\ref{fig1} and~\ref{fig3}) or localized at  $\tilde{\eta}>\tilde{\eta}_{\rm LDT}$ (Fig.~\ref{fig4}). A distinctive feature of the cavity system is that each mode within a given family $\mathcal{N}(\mu)$ corresponds to a different number of cavity photons, as it is expressed by $A(\mathcal{N})$. Consequently, the situation, which is typical of conservative systems, where several stable nonlinear modes may coexist for the same $\mathcal{N}$, manifests here as bi- or multistability, i.e., the existence of different numbers of photons for identical cavity parameters (pumping strength, decay rate) and detunings from resonance.

A bistable regime in a cavity with deep symmetric ($\vartheta=0$ in our notations) incommensurate lattice for blue atomic detuning was reported in earlier studies~\cite{Zhou2011}. The stable states were localized and extended ones. Meanwhile, the latter ones strongly depend on the boundary conditions, i.e. somewhat artificial for systems subject to periodic or zero Dirichlet boundary conditions (this constraint can be avoided by using quasi-periodic boundary conditions~\cite{Gao2025}). At the same time, at $\vartheta=0$, authentic localized states, except possibly the ground state localized at the origin, which are consistent with the parity symmetry of the potential are two-hump ones~\cite{Prates2022}. The results presented here reveal the generality of the bistability phenomenon, which can occur in relatively shallow quasi-periodic potentials, manifest itself in different forms, and involves two localized states (i.e., is not affected by the choice of the boundary conditions). Furthermore, two types of bi- (or multi-) stable states can be distinguished. 
  
The first type is {\em multi-stability} arising from multiple possible distributions of the BEC corresponding to different families of solutions, which is typical for conservative nonlinear systems.  For a given family at fixed $\cN$ there corresponds a certain number of photons $A$, hence considering modes belonging to different families one obtains different atomic distributions coupled with different numbers of photons. This type of bistability can occur for all sings of the detunings. An example is provided by modes A and B shown in Fig.~\ref{fig2} and marked by the empty circles in Figs.~\ref{fig1} (b) and (c). A continuous transition between such states in the parameter space is not possible.

Irrespective of the signs of the detunings, intersections between two families can be identified (they are indicated by crosses in the figures). At these points a pseudo-degeneracy occurs: two distinct atomic-density distributions correspond to the same number of atoms and the same cavity-field amplitude.

The second type of bistability refers to multivalued dependence $A(\mathcal{N})$ for a single family which is observed in the case of blue detunings $\Delta_a,\Delta_c>0$ for sufficiently large number of atoms. In this case, typical for systems with hysteresis behavior, like the one considered in~\cite{Zhou2011}, bistable states belong to the same family [see Figs.~\ref{fig1}(a) and (b)]. In the parameter space, these states are (formally) connected through a continuous change of the system parameters.

\section{On dynamics of localized states}
\label{sec:dynamics}

\begin{figure}
	\includegraphics[width=\columnwidth]{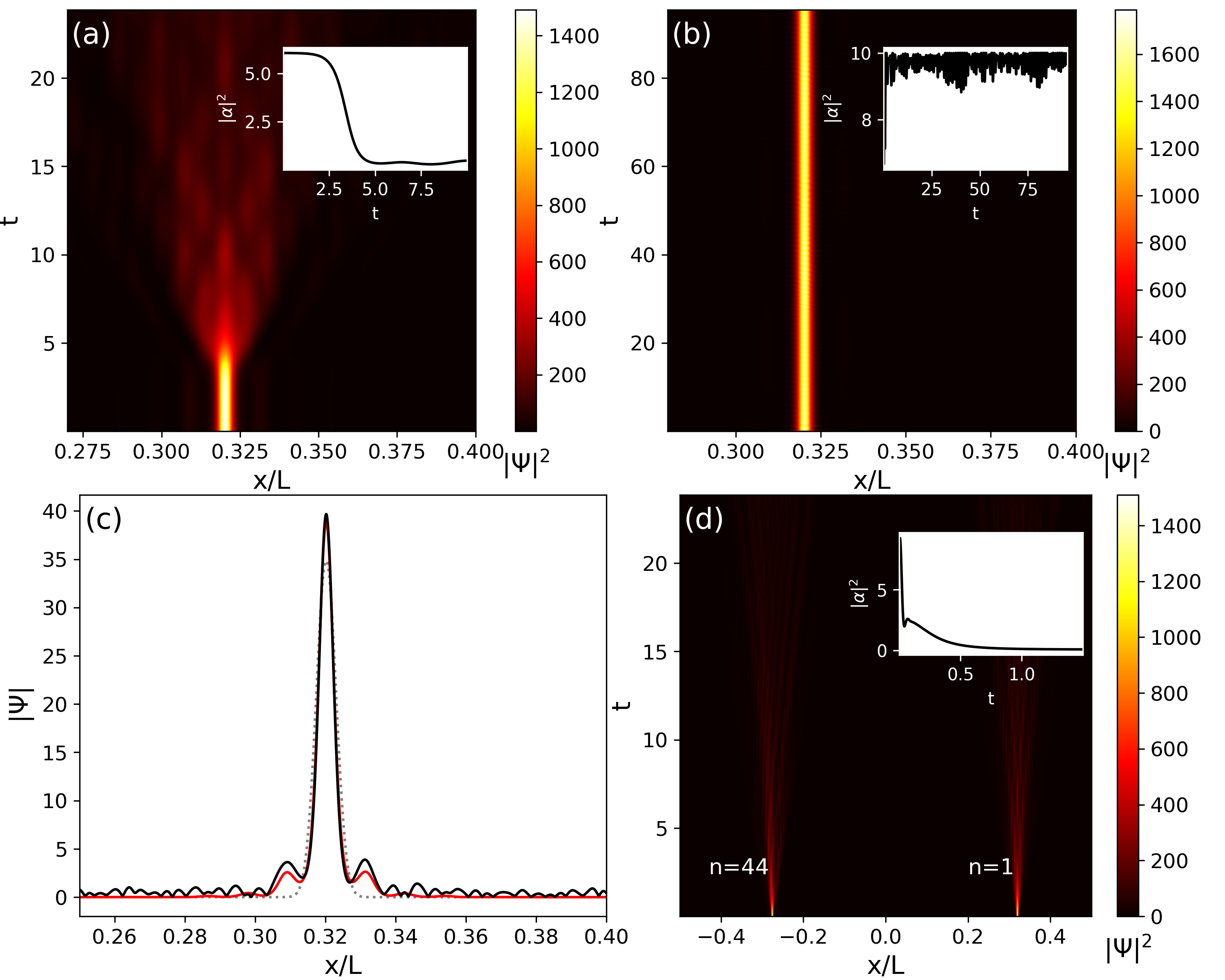}
	\caption{(a) Evolution of the (numerically) exact unstable state C (see Fig.~\ref{fig1}), for $\sigma=+1$, $\Delta=150$ and $\tilde\eta=141$ ($\eta\approx200$),  with $\alpha(0)=0.995\alpha_{\rm st}$, where $\alpha_{\rm st}\approx1.92-1.55i$ ($A\approx3.03$) is the amplitude of the cavity mode corresponding to C, and to $\cN\approx 990$. (b) Evolution of a Gaussian wave-packet (\ref{gaussian}), with the center of mass and variance equal to those of the mode D (see Fig. \ref{fig1}(b)), excited with an initial cavity amplitude equal to the stationary value $\alpha_{\rm st}$ corresponding to mode D. (c) The resulting spatial profile at the end of simulation  at $t\approx 95$ (black solid line), over the target mode D (red solid line), and the initial Gaussian spatial profile (gray dashed line). (d) Two stationary modes, from families $n=1$ and $n=44$, with the same stationary cavity amplitude, $\alpha_{\rm st}$, are excited at the same time and for an initial $\alpha(t=0)=\alpha_{\rm st}$.}
	\label{fig5}
\end{figure}

Now we address several aspects of the dynamics of localized states obtained through direct numerical integration of Eqs.~(\ref{eq:Psi}), (\ref{eq:alpha}). Starting with the evolution of unstable modes, Fig.~\ref{fig5} (a) illustrates a representative example of decay where state C, indicated by a circle in Fig.~\ref{fig1}(b), is chosen and the cavity field is initially perturbed as $\alpha(0)=0.995\,\alpha_{st}$. Owing to the coupling between the field and the atoms, this perturbation also affects the corresponding stationary atomic distribution $\psi(x)$. The perturbed state C disperses after $t\approx 5$ ($\sim 0.1\,$ms in physical units), evolving towards extended states. The dispersion continues until atomic density reaches the boundaries of simulation domain (i.e. $|x|=L/2$), beyond which the modeling using a finite interval is no longer applicable.  

Excitation of localized modes in a cavity is a challenging problem because of both complex shape of atomic distributions and long-range interactions inherent for such systems. However, judicious choice of the initial data allows one to solve this problem. Excitation of stable state D [see Fig.~\ref{fig1}(a)] is illustrated in Fig.~\ref{fig5}(b), where we show the evolution, and in Fig.~\ref{fig5}(c), where we show the initial state (dotted line), the exact mode (solid red line) and the atomic state at $t=103$ (solid dark line). To excite this mode we used the Gaussian initial condition  
\begin{align}
	\label{gaussian}
	\Psi(x,t=0)=\frac{N_{\rm D}^{1/2}}{(2\pi \sigma^2_{\rm D})^{1/4}}
	 e^{-\frac{(x-X_{\rm D})^2}{4\sigma^2_{\rm D}} }
\end{align}
where  $X_{\rm D}\approx89.5$, $\sigma^2_{\rm D}\approx 0.311$, and $N_{\rm D}\approx1697$  denote, respectively, the c.m., the variance, and $105\%$ of the number of atom number corresponding to the exact mode D. The initial amplitude of the cavity field was chosen as the respective $\alpha_{\rm st}$ which in the case at hand was $\alpha_{\rm st}\approx2.85+0.95i$.
    
Once the Gaussian distribution is excited within the cavity, it rapidly adapts to the profile of mode D. At the same time a residual fraction of atoms spreads throughout the cavity. Their presence, which is guaranteed by the conservation of total number of atoms, leads to a weakly oscillating photon number at a frequency $\approx 50$ kHz in physical units, as illustrated in the inset in Fig.~\ref{fig5}(b).
 
As the third dynamical effect, Fig.~\ref{fig5}(d) presents the evolution initiated by the simultaneous excitation of two spatially separated modes, each of which is stable when excited individually. The chosen modes $\phi_1$ and $\phi_{44}$ belong to different families, $n=1$ and $n=44$, with $N_1\approx 1617$ ($\cN_1\approx808$) and $N_{44}\approx1513$ ($\cN_{44}\approx757$), i.e., the total number of atoms in the cavity is $N= N_1+N_2\approx3130$. It is important to note, however, that when these states are excited individually, they both correspond to the same cavity-field amplitude $\alpha_{\rm st}$ ($A\approx4.5$). Thus, the initial wave-function in simulations shown in Fig.~\ref{fig5} (d) is chosen as $\Psi(x,t=0)=\sqrt{N_1}\phi_{1}(x)+\sqrt{N_{44}}\phi_{44}(x)$ and  $\alpha(t=0)=\alpha_{\rm st}$ computed from (\ref{A}).

We observe that, although the two modes are individually stable, spatially well separated, and correspond to the same intensity of the cavity mode, they cannot coexist in the same cavity due to the long-range interactions mediated by cavity photons. As a result, the modes excited simultaneously are rapidly destroyed with time, and the number of photons (shown in the inset) decays from its initial value of $|\alpha|^2\approx9$ to $|\alpha|^2\approx0.4$. This behavior demonstrates that the system exhibits the characteristics of a logical XOR gate. Indeed, by assigning a logical “true” (1) to the presence of a localized mode and “false” (0) to its absence, the output is true only for the inputs $|0,1\rangle$, $|1,0\rangle$, while it is false for the inputs $|0,0\rangle$, $|1,1\rangle$. While this emulation of a logic gate was carried out for a condensate, previous observations of single-atom localization~\cite{Rojan2016,Major2018} suggest that the proposed scheme should also be applicable in the regime of a single (or a few) atoms.

\section{Conclusions}

In this work we described atomic BECs interacting with cavity photons which combine features of both conservative and dissipative systems, remaining inherently nonlinear even in the absence of inter-atomic interactions. When an external lattice potential with a period incommensurate with the back-action potential is introduced, localized atomic states coupled to cavity modes can exist within certain ranges of the chemical potential, and respectively withing certain range of photon numbers. Considering relatively shallow one-dimensional potentials, where the tight-binding approximation is not applicable, we have identified families of localized modes for different signs of the detuning between probe field and cavity or atomic resonances. The stability (or instability) of these nonlinear modes were confirmed by direct numerical propagation. Two distinct types of bistability, involving localized modes and corresponding to the existence of different cavity field amplitudes for identical system parameters (including atom number),  were found. The first type arises from the coexistence of multiple families of nonlinear modes, while the second results from the multivalued dependence of individual families on system parameters. We also have shown that the long-range interactions mediated by the cavity field enable regimes where the system functions as a logical gate, with localized modes representing logical truth values. To conclude, while in this work we have focused on the simplest one-dimensional case, extensions to more complicated three-dimensional models remain open, where other approaches, such as the potential harmonics expansion~\cite{Ripelle1983,Das2004}, may be of use.

\acknowledgments

The work was supported by the Fundação para a Ciência e Tecnologia under the project 2023.13176.PEX (DOI https://doi.org/10.54499/2023.13176.PEX) and by national funds, under the Unit CFTC - Centro de Física Teórica e Computacional, reference UID/00618/2023, financing period 2025-2029.




\end{document}